\pdfoutput=1

\documentclass[useAMS,usenatbib]{mnras}

\usepackage{graphicx,amssymb}
\usepackage[normalem]{ulem}
\usepackage[usenames,dvipsnames]{xcolor}
\usepackage{soul}

\title[Stability of particles in WD J0914+1914]
{Short-term stability of particles in the WD J0914+1914 white dwarf planetary system}

\author[]{Euaggelos E. Zotos$^1$, Dimitri Veras$^{2,3}$\thanks{E-mail: d.veras@warwick.ac.uk}\thanks{STFC Ernest Rutherford Fellow}, Tareq Saeed$^4$, Luciano A. Darriba$^{5,6}$
\\
$^{1}$Department of Physics, School of Science, Aristotle University of Thessaloniki, 541 24, Thessaloniki, Greece
\\
$^{2}$Centre for Exoplanets and Habitability, University of Warwick, Coventry CV4 7AL, UK
\\
$^{3}$Department of Physics, University of Warwick, Coventry CV4 7AL, UK
\\
$^{4}$Nonlinear Analysis and Applied Mathematics (NAAM)-Research Group, Department of Mathematics, Faculty of Science, \\
King Abdulaziz University, P.O. Box 80203, Jeddah 21589, Saudi Arabia
\\
$^{5}$Instituto de Astrof\'{\i}sica de La Plata, CCT La Plata-CONICET-UNLP Paseo del Bosque S/N (1900), La Plata, Argentina
\\
$^{6}$Facultad de Ciencias Astron\'omicas y Geof\'\i sicas, Universidad Nacional de La Plata Paseo del Bosque S/N (1900), La Plata, Argentina
}

\begin{document}
\label{firstpage}
\pagerange{\pageref{firstpage}--\pageref{lastpage}}
\maketitle

\begin{abstract}
Nearly all known white dwarf planetary systems contain detectable rocky debris in the stellar photosphere. A glaring exception is the young and still evolving white dwarf WD J0914+1914, which instead harbours a giant planet and a disc of pure gas. The stability boundaries of this disc and the future prospects for this white dwarf to be polluted with rocks depend upon the mass and orbit of the planet, which are only weakly constrained. Here we combine an ensemble of plausible planet orbits and masses to determine where observers should currently expect to find the outer boundary of the gas disc. We do so by performing a sweep of the entire plausible phase space with short-term numerical integrations. We also demonstrate that particle-star collisional trajectories, which would lead to the (unseen) signature of rocky metal pollution, occupy only a small fraction of the phase space, mostly limited to particle eccentricities above 0.75. Our analysis reveals that a highly inflated planet on a near-circular orbit is the type of planet which is most consistent with the current observations.
\end{abstract}


\section{Introduction}

A common motivation for dynamical analyses of main-sequence planetary systems is to identify locations where a major planet remains stable. However, the focus in white dwarf planetary systems is markedly different: to identify instability, and specifically of the type where minor planets will collide with the star.

This difference in focus arises due to the observables in both types of systems. In almost every known white dwarf planetary system, the stellar photosphere contains rocky debris \citep{zucetal2003,zucetal2010,koeetal2014,couetal2019}. Because the extent of the chemical information encoded in this debris is unprecedented within exoplanetary science \citep{ganetal2012,faretal2013,juryou2014,xuetal2017,haretal2018,holetal2018,doyetal2019,swaetal2019,bonetal2020}, it provides a unique window into the composition of exoplanetary material.

Identifying the dynamical history of this debris then allows one to link its formation location \citep{haretal2018} with its subsequent evolution as the star traverses its post-main-sequence phases \citep{veras2016}. Eventually, this material, which can survive on au-scales for billions of years \citep{verhen2020}, is gravitationally perturbed towards and accreted by the white dwarf. The perturbation process has been investigated extensively \citep{bonetal2011,debetal2012,frehan2014,bonver2015,antver2016,antver2019,hampor2016,petmun2017,steetal2017,steetal2018,musetal2018,smaetal2018}, but only with major planets at au-scale distances or with companion stars.

This au-scale assumption about where major planets should orbit a white dwarf is well-founded because planets on tight orbits are engulfed by the star during the giant branch phases \citep{kunetal2011,musvil2012,adablo2013,norspi2013,viletal2014,madetal2016,staetal2016,galetal2017,raoetal2018,sunetal2018}. No major planets should exist within about 1-2 au of a white dwarf unless they are perturbed there by other major planets  \citep{debsig2002,veretal2013,veretal2016,veretal2018,voyetal2013,musetal2014,vergan2015,ronetal2020}. Indeed, until 2019, no major planets in such tight orbits were detected, despite the discoveries of several minor planets \citep{vanetal2015,manetal2019,vanetal2019}.

Finally, in 2019, \cite{ganetal2019} reported the discovery of a giant planet -- and specifically an ice giant planet -- orbiting white dwarf WD J0914+1914 at an approximate distance of just 0.07 au ($15R_{\odot}$). This system is unique not only because of the presence of a major planet on such a close orbit but also because the white dwarf photosphere does not contain any rocky debris. Instead, the volatile species detected in the stellar photosphere arise from the evaporation of the planet's atmosphere. These species also form a disc of gas located in the approximate range of  0.0046-0.046 au ($1-10R_{\odot}$). For perspective, the radius of the white dwarf itself is just $\sim 10^{-2} R_{\odot}$.

Both the extent of the gas disc as well as the lack of rocky debris motivate dynamical questions about the role of the planet, denoted as WD J0914+1914~b, in shaping the system.  \cite{veras2020} found that WD J0914+1914~b acts as an effective barricade for pebbles and most boulders which are {\it radiatively} dragged towards the white dwarf. However, he only sampled parameter space which was relevant to those bodies and considered only radiative drag and one type of planet. In fact, the mass and orbit of WD J0914+1914~b are weakly constrained from the observations \citep{ganetal2019}. Fortunately, theoretical considerations about tidal circularization of planets perturbed towards white dwarfs can place stricter constraints \citep{verful2019,veretal2019,ocolai2020}.

Because WD J0914+1914 has existed as a white dwarf for only about 13 Myr, WD J0914+1914~b could only reach a separation of 0.07 au through a combination of a gravitational scattering event (suggesting additional major planets in the system) followed by quick chaotic tidal interactions \citep{verful2020}. These interactions are a function of the mass and radius of WD J0914+1914~b. As a result, there is a degeneracy with respect to the planet's physical and orbital properties.

Here, we take into account this degeneracy while examining the phase space structure of particle orbits in the immediate vicinity of the planet. To enable a broad exploration, we perform full but quick integrations of three-body systems and determine which orbits are bounded (section \ref{clas}).  Then, in Section \ref{simul} we establish and justify the range of parameters that we simulate. We report and describe the results in Section \ref{res}, discuss them in Section \ref{disc} and conclude in Section \ref{sum}.

\section{Integrator and orbit types}
\label{clas}

For modelling this exoplanetary system, we will integrate the equations of motion for the general three-body problem. In our case, the first body is the white dwarf (with mass $M_{\star}$), the second body is the exoplanet (with mass $M_{\rm pl}$), and the third body is a test particle (with mass $M$). For simplicity, we adopt a heliocentric (astrocentric), non-inertial reference system, in which the primary body (white dwarf) is located at the origin of the coordinates $O(0,0,0)$ and its position is fixed. On the other hand, both the exoplanet and the test particle are free to move, with respective position vectors $\vec{r}_{\rm pl}$ and $\vec{r}$. Thus, the motion of the exoplanet, relative to the white dwarf, is given by
\begin{equation}
\ddot{\vec{r}}_{\rm pl} = - G \left(M_{\star} + M_{\rm pl} \right) \frac{\vec{r}_{\rm pl}}{|\vec{r}_{\rm pl}|^3} + G M \left(\frac{\vec{r} - \vec{r}_{\rm pl}}{|\vec{r} - \vec{r}_{\rm pl}|^3} - \frac{\vec{r}}{|\vec{r}|^3} \right),
\end{equation}
while the motion of the test particle, relative to the white dwarf, is given by
\begin{equation}
\ddot{\vec{r}} = - G \left(M_{\star} + M \right) \frac{\vec{r}}{|\vec{r}|^3} + G M_{\rm pl} \left(\frac{\vec{r}_{\rm pl} - \vec{r}}{|\vec{r}_{\rm pl} - \vec{r}|^3} - \frac{\vec{r}_{\rm pl}}{|\vec{r}_{\rm pl}|^3} \right).
\end{equation}

Given our choice of reference frame, the integrator requires initial position and velocity vectors for only the planet and test particle. We provide these through the common orbital elements of semimajor axis $a$, eccentricity $e$, inclination $i$, argument of pericentre $\omega$, longitude of ascending node $\Omega$ and mean anomaly $\mathcal{M}$. Also we assume that the mass $M$ of the test particle is significantly smaller than the masses of the white dwarf and the planet. Therefore, in our computations, we set $M = 0$.

For our computations, we adopt a system of units where $G = k^2$, with $k = 0.01720209895$ being the Gaussian gravitational constant. Then, within our integrator, our unit of time is days, our unit of length is au, and our unit of mass is Solar masses. However, because our results are specifically applicable to the WD~J0914+1914 system, we report all of our results in Section \ref{res} in physical units.

For the numerical integration of the equations of motion a double precision Bulirsch-Stoer \verb!FORTRAN 77! algorithm \citep{PTVF92} was used. Throughout our calculations, the numerical errors, related to the values of the total orbital energy and the total angular momentum of the system, were of the order of $10^{-12}$ (or smaller), thus indicating a sufficient conservation of both quantities.

\begin{figure*}
\centerline{\bf \LARGE \underline{Orbit classification within integrator}}
\includegraphics[width=20cm]{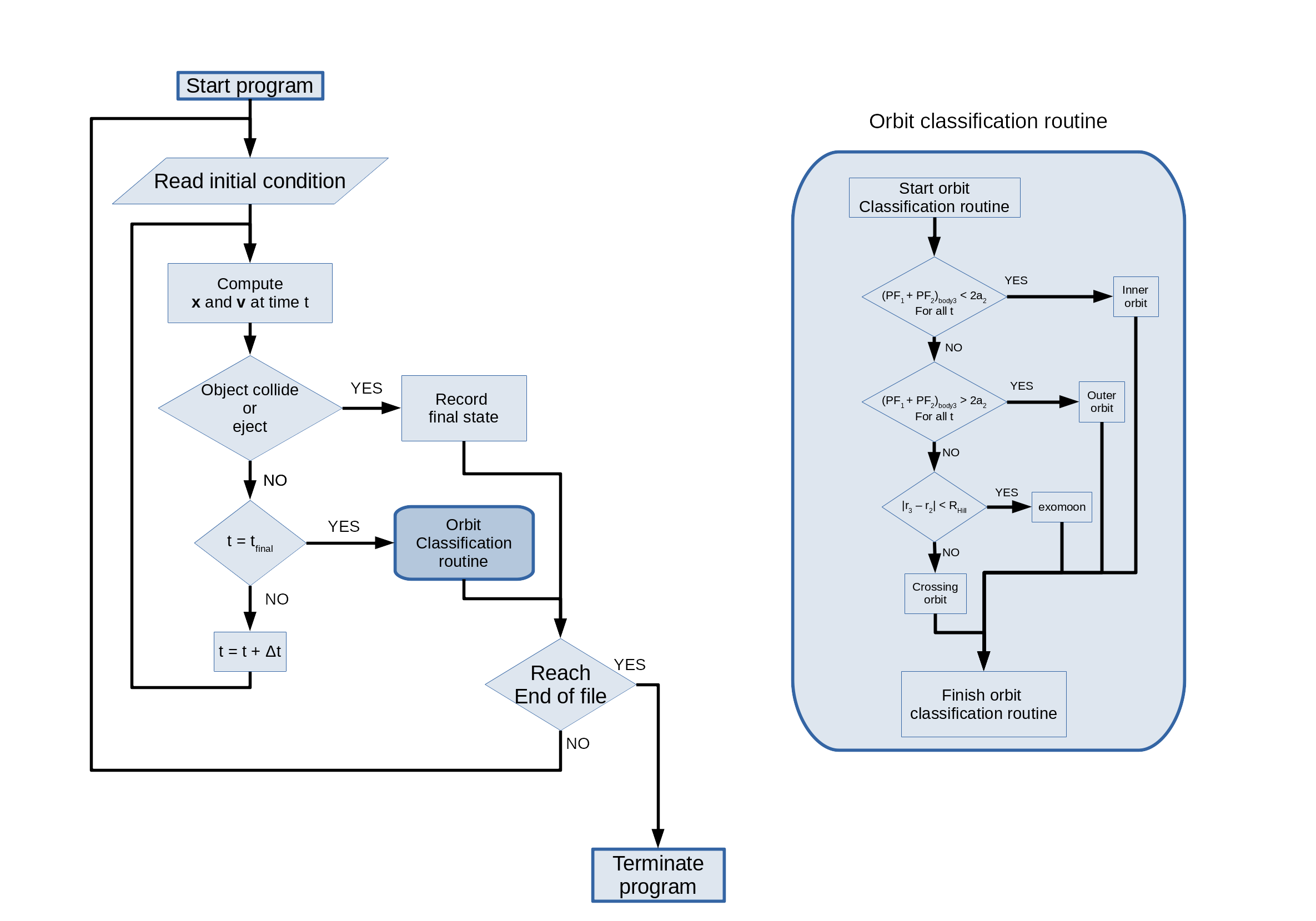}
\caption{
How orbits are classified within our integrator. The left chart corresponds to the general program flow. The right chart shows the details of the inside of the orbit classification routine, which is highlighted in the left chart in the bubble with a darker tone. Because the test particle's orbit is an ellipse, the expression $PF_i$ indicates the distance from the test particle's position on its orbit to each of the orbit's foci $i$.
}
\label{flow}
\end{figure*}

\begin{figure*}
\centerline{\bf \LARGE \underline{Orbit classification for stable orbits}}
\includegraphics[width=16cm]{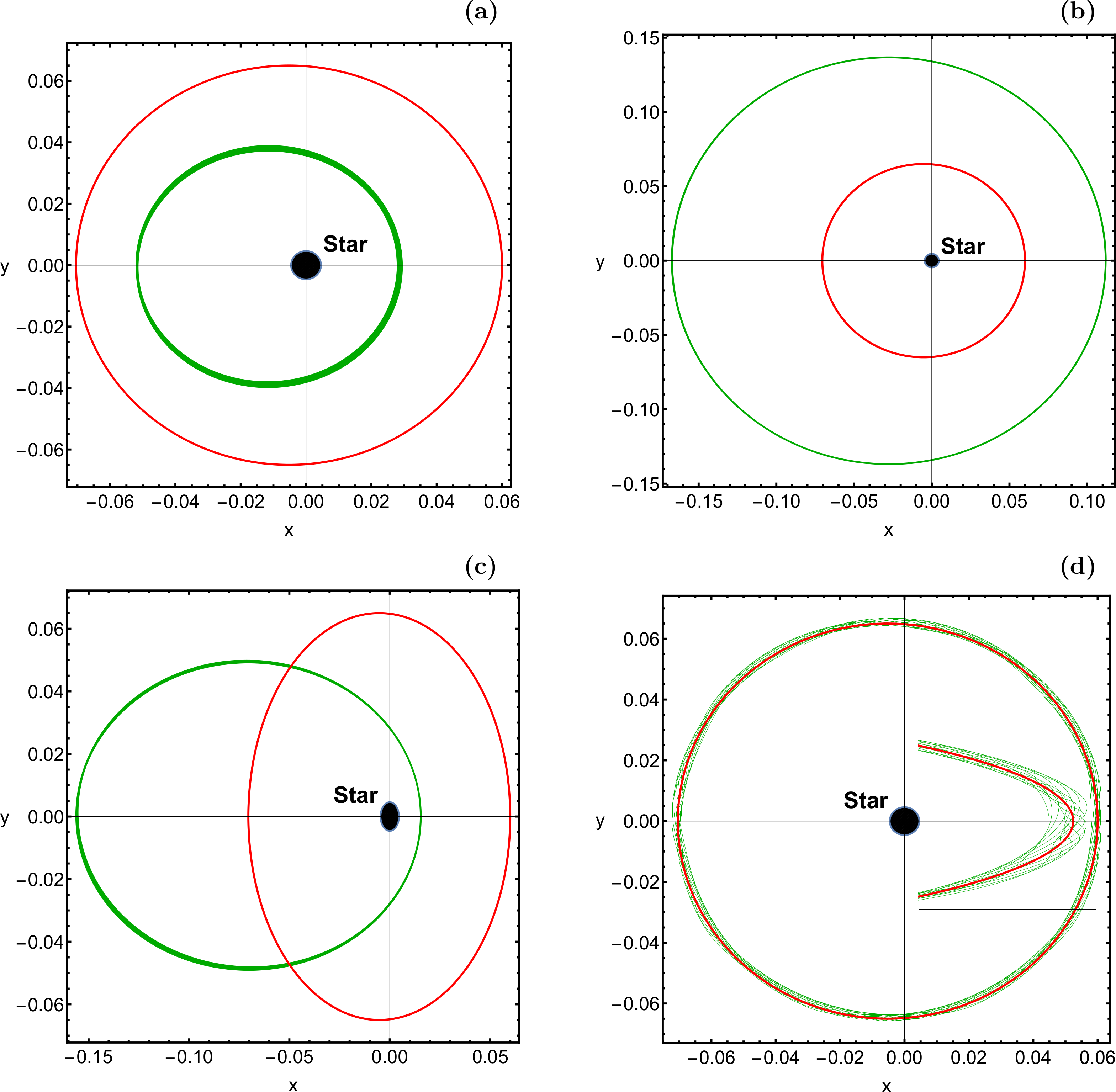}
\caption{
Schematic examples of the outcomes of our orbit classification for stable orbits (the physical units of $x$ and $y$ are unimportant). The green orbits are the particle's orbits, and the red orbits are the planet-star's orbits. The panels respectively show {\bf a:} a circumstellar ``inner'' orbit, {\bf b:} a circumbinary ``outer'' orbit, {\bf c:} a crossing orbit, and {\bf d:} a circumplanetary ``moon'' orbit.
}
\label{schem}
\end{figure*}

\begin{figure*}
\centerline{\bf \LARGE \underline{Orbit classification for unstable orbits}}
\includegraphics[width=16cm]{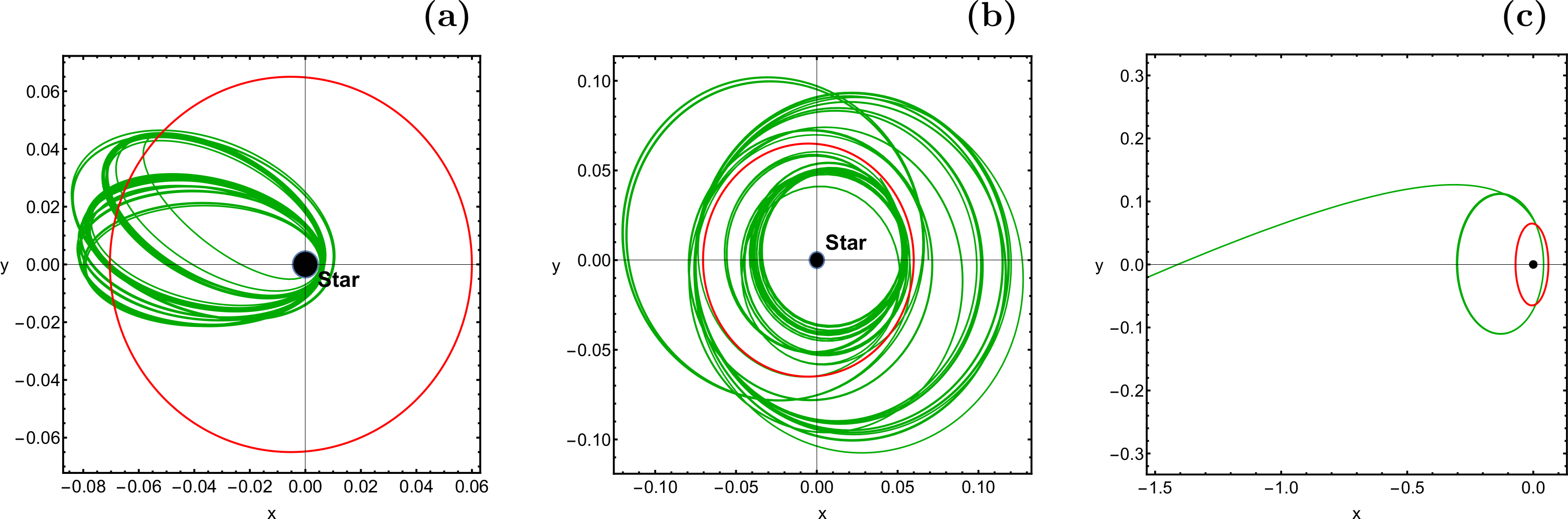}
\caption{
Idem as Fig. \ref{schem} for unstable orbits. The color-code is the same as Fig. \ref{schem}. The panels show {\bf a:} a collision with the star, {\bf b:} a collision with the planet, and {\bf c:} an escape after of the system at about 3000 time units
}
\label{orbs}
\end{figure*}

Within our integrator, we have adopted an algorithm to determine which particle orbits are bounded and in what manner (the planet's orbit is unperturbed). We lay out our scheme in the flowchart in Fig. \ref{flow} and illustrate schematically the different types of orbits in Figs. \ref{schem}-\ref{orbs}. We describe each orbit type as follows.

\begin{itemize}

\item {\bf Circumstellar orbit, or ``inner orbit''}. The particle orbits the star entirely within the planet-star orbit, throughout the evolution (see panel {\bf a} of Fig.~\ref{schem}).

\item {\bf Circumbinary orbit, or ``outer orbit''}. The particle orbits both the star and the planet external to the planet-star orbit, throughout the evolution (see panel {\bf b} of Fig.~\ref{schem}).

\item {\bf Crossing orbit}. Both the particle and the star-planet orbits intersect, but the particle remains stable throughout the evolution (see panel {\bf c} of Fig.~\ref{schem}).

\item {\bf Circumplanetary orbit, or ``exomoon orbit''}. Particle orbits the planet just as a moon would, throughout the evolution (see panel {\bf d} of Fig.~\ref{schem}).

\item {\bf Stellar polluting orbit, or ``star collision orbit''}. The particle collides with the star, polluting the white dwarf with metals (see panel {\bf a} of Fig.~\ref{orbs}).

\item {\bf Planetary collision orbit, or ``planet collision orbit''}. The particle collides with the planet (see panel {\bf b} of Fig.~\ref{orbs}).

\item {\bf Escaping orbit, or ``escape orbit''}. The particle escapes the planetary system (see panel {\bf c} of Fig.~\ref{orbs}).

\end{itemize}

Each integration is run for $10^4$ orbital periods of the planet. Although the actual time to which this value corresponds depends on the adopted semimajor axis of the planet, the minimum duration of any of our integrations is a few hundred years, well exceeding immediately observable time frames. The list of orbits above is complete in the sense that at the end of the integrations, the particle orbit can always be classified according to one of those seven types.

No simple, uniform recipe (explicit formulae) for the secular timescales of our setup exists because these timescales depend on particular effects, such as the true secular resonance in high-eccentricity regime. Similar to mean-motion resonance time-scales, the secular timescales can be determined by the characteristic frequency (period) of the given effect multiplied by some number of perturbation cycles (usually a few thousands) depending on the perturbation strength.

\section{Simulation parameters}
\label{simul}

In this section we describe the parameter ranges of our simulations. In what follows, we will present color-coded basin diagrams containing $500 \times 500$ grids of initial conditions. These grids are coloured according to the final state of the test particle.

\subsection{The star}

We start with the star, and assume that WD J0914+1914 has a mass of $M_{\star} = 0.56M_{\odot}$ \citep{ganetal2019}. We don't need to define its physical radius because, for collision detection, we instead adopt its disruption, or Roche radius, $R_{\rm Roche}$. The value of $R_{\rm Roche}$ can vary by a factor of several depending on the physical and orbital properties of the disrupting object \citep{veretal2017}. Because such a variation is likely to have a negligible effect on the outcome, for computational expediency we adopt $R_{\rm Roche} = R_{\odot}$ for all simulations.

\subsection{The planet}

Regarding the planet, \cite{verful2020} placed coupled constraints on its mass $M_{\rm pl}$, radius $R_{\rm pl}$ and orbit, which is partly characterized with semimajor axis $a_{\rm pl}$, eccentricity $e_{\rm pl}$ and orbital pericentre $q_{\rm pl} = a_{\rm pl} - a_{\rm pl}e_{\rm pl}$. Furthermore, if the planet underwent thermalization events during chaotic tidal evolution \citep{verful2019}, then it could have self-disrupted, leaving behind an arbitrarily small core. We hence adopt a wide range of planet masses, from $M_{\oplus}$ to $M_{\rm Saturn}$.

The planet's radius is then dictated by our adopted density. This density crucially determines tidal migration and circularization timescales, which are in turn constrained by the cooling age of the star (13 Myr). \cite{verful2020} found that, if the planet currently resides on a circular orbit ($e_{\rm pl} = 0$), then it would need to represent a ``Super-puff'' (term from \citealt*{leechi2016}), which is a highly inflated planet with a low density ($\approx 0.1$ g/cm$^3$). Higher density planets are allowed for $e_{\rm pl} > 0$. Further, the current value of $e_{\rm pl}$ further dictates the value of $q_{\rm pl}$ through conservation of angular momentum over time.

In summary, we adopt three different planet types and eight different combinations of masses, radii and orbits. For ease of reference, we categorize these combinations with the following cases:

\begin{itemize}

\item Case I: (Super-puff)

$M_{\rm pl} = 1M_{\oplus}$, $R_{\rm pl} = 3.78R_{\oplus}$, $q_{\rm pl} = a_{\rm pl} = 0.07$ au ($e_{\rm pl}=0.0$)

\

\item Case II: (Super-puff)

$M_{\rm pl} = 4M_{\oplus}$, $R_{\rm pl} = 6R_{\oplus}$, $q_{\rm pl} = a_{\rm pl} = 0.07$ au ($e_{\rm pl}=0.0$)

\

\item Case III: (Neptune)

$M_{\rm pl} = 1M_{\rm Neptune}$,  $R_{\rm pl} = 1R_{\rm Neptune}$, $q_{\rm pl} = 0.046$ au, $a_{\rm pl} = 0.05$ au ($e_{\rm pl}=0.08$)

\

\item Case IV: (Neptune)

$M_{\rm pl} = 1M_{\rm Neptune}$,  $R_{\rm pl} = 1R_{\rm Neptune}$, $q_{\rm pl} = 0.035$ au, $a_{\rm pl} = 0.070$ au ($e_{\rm pl}=0.50$)

\

\item Case V: (Neptune)

$M_{\rm pl} = 1M_{\rm Neptune}$,  $R_{\rm pl} = 1R_{\rm Neptune}$, $q_{\rm pl} = 0.023$ au, $a_{\rm pl} = 0.46$ au ($e_{\rm pl}=0.95$)

\

\item Case VI: (Saturn)

$M_{\rm pl} = 1M_{\rm Saturn}$,  $R_{\rm pl} = 1R_{\rm Saturn}$, $q_{\rm pl} = 0.06$ au, $a_{\rm pl} = 0.0652$ au ($e_{\rm pl}=0.08$)

\

\item Case VII: (Saturn)

$M_{\rm pl} = 1M_{\rm Saturn}$,  $R_{\rm pl} = 1R_{\rm Saturn}$, $q_{\rm pl} = 0.045$ au, $a_{\rm pl} = 0.090$ au ($e_{\rm pl}=0.50$)

\

\item Case VIII: (Saturn)

$M_{\rm pl} = 1M_{\rm Saturn}$,  $R_{\rm pl} = 1R_{\rm Saturn}$, $q_{\rm pl} = 0.03$ au, $a_{\rm pl} = 0.6$ au ($e_{\rm pl}=0.95$)

\end{itemize}

The justification for these scenarios is as follows. Cases I and II sample the fiducial ``Super-puff'' circular case, but with different masses. Cases III, IV and V, instead, consider a Neptune-like planet. According to Fig. 1 of \cite{verful2020}, tidal circularization in such a planet would have been triggered in the pericentre range $(0.013 \ {\rm au}, 0.023 \ {\rm au})$, meaning that its current pericentre due to angular momentum conservation is similar for $e_{\rm pl}=0.95$, a factor of 1.5 higher for $e_{\rm pl}=0.5$ $(0.020 \ {\rm au}, 0.035 \ {\rm au})$ and a factor of about 2.0 higher for $e_{\rm pl}=0.08$ $(0.026 \ {\rm au}, 0.046 \ {\rm au})$. We adopt the upper ends of these ranges. Finally, cases VI, VII and VIII consider a Saturn-like planet with the same three eccentricities. For this type of planets, tidal circularization would have originally been triggered in the pericentre range $(0.02 \ {\rm au}, 0.03 \ {\rm au})$.

In the general three-body problem, the planet's orbit is also defined through its inclination $i_{\rm pl}$, argument of pericentre $\omega_{\rm pl}$, and longitude of ascending node $\Omega_{\rm pl}$. We set these variables to $0^{\circ}$ throughout. Doing so, it establishes the reference plane and orientation of the orbit. We also begin all integrations with the planet's mean anomaly $\mathcal{M}_{\rm pl} = 0^{\circ}$, such that the planet initially resides at the orbit's pericentre (we will vary the initial mean anomaly of the particle).

Because one outcome of our simulations will be circumplanetary orbits, we also define the planet's Hill radius, $R_{\rm Hill}$. We define a circumplanetary orbit as one where the particle remains within the Hill radius for the duration of the simulation.  The expression for $R_{\rm Hill}$ is dependent on both $a_{\rm pl}$ and $e_{\rm pl}$. We choose to use the planet's pericentre value by modifying Eq. (B5) of \cite{peawya2014} to:

\begin{equation}
R_{\rm Hill} = a_{\rm pl} \left(1 - e_{\rm pl}\right)
                       \left[ \frac{M_{\rm pl}}{\left(3 + e_{\rm pl}\right) M_{\star}} \right]^{1/3}.
\end{equation}

\noindent{}Therefore, the value of $R_{\rm Hill}$ is, hence, different for each of our eight cases.

\subsection{The particles}

For each case, we will perform a simulation suite exploring the parameters space of particle orbits. We first must decide which particle parameters are most relevant to explore. Denote the initial orbital elements of the particle without subscripts: $\left(a,e,i,\omega,\Omega\right)$. Only for the particle's initial mean anomaly, we explicitly use a subscript ($\mathcal{M}_0$).

Although the particles may represent constituents of the gas disc (which ranges from 0.0046 au - 0.046 au), they could also represent rocky boulders external to the gas disc (or any other object which could be treated as a test particle). In principle, there is no restriction on $e$ or $i$. For gas disc particles, $e \approx i \approx 0$ is a reasonable assumption. Instead, rocky particles external to the planet could have any orbital eccentricity or inclination. These values may or may not have been radiatively damped through the relatively high luminosity ($\approx 0.1L_{\odot}$) of this white dwarf \citep{veras2020} and could be at any stage of damping.

In all cases, we varied $a$ along the $x$-axis of our two-dimensional ``basin'' diagrams. The chosen ranges encompass the radial width of the planet's orbit. Along the $y$-axis, we chose three variables to explore: the initial mean anomaly of the particle $\mathcal{M}_0$, as well as $e$ (for both $\omega = 0^{\circ}$ and $\omega =180^{\circ}$) and $i$.

\section{Simulation results}
\label{res}

\begin{figure*}
\centerline{\bf \LARGE \underline{Circular Super-puff planets}}
\includegraphics[width=16cm]{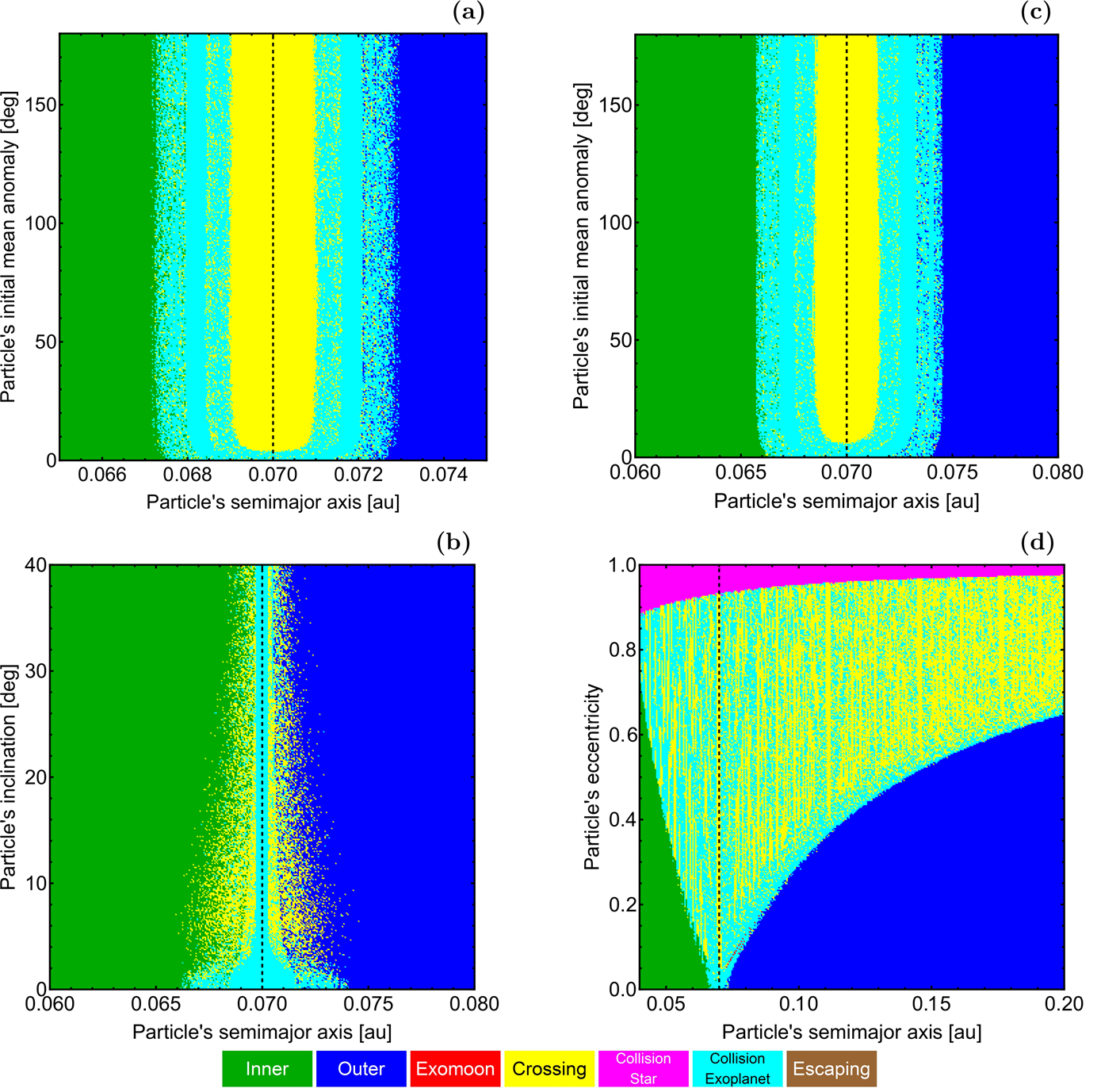}
\caption{
Particle orbit classification in the WD~J0914+1914 system, assuming planet properties corresponding to Case I (panel {\bf a}) and Case II (panels {\bf b}, {\bf c}, {\bf d}). Particle orbital variables which are not varied in a particular plot are set to zero. The plots show that the planet would not disturb a circular gas disc, and only pollutes the white dwarf when $e > 0.75$.
}
\label{puff}
\end{figure*}

\begin{figure*}
\centerline{\bf \LARGE \underline{Eccentric Neptune-like planets}}
\includegraphics[width=14cm]{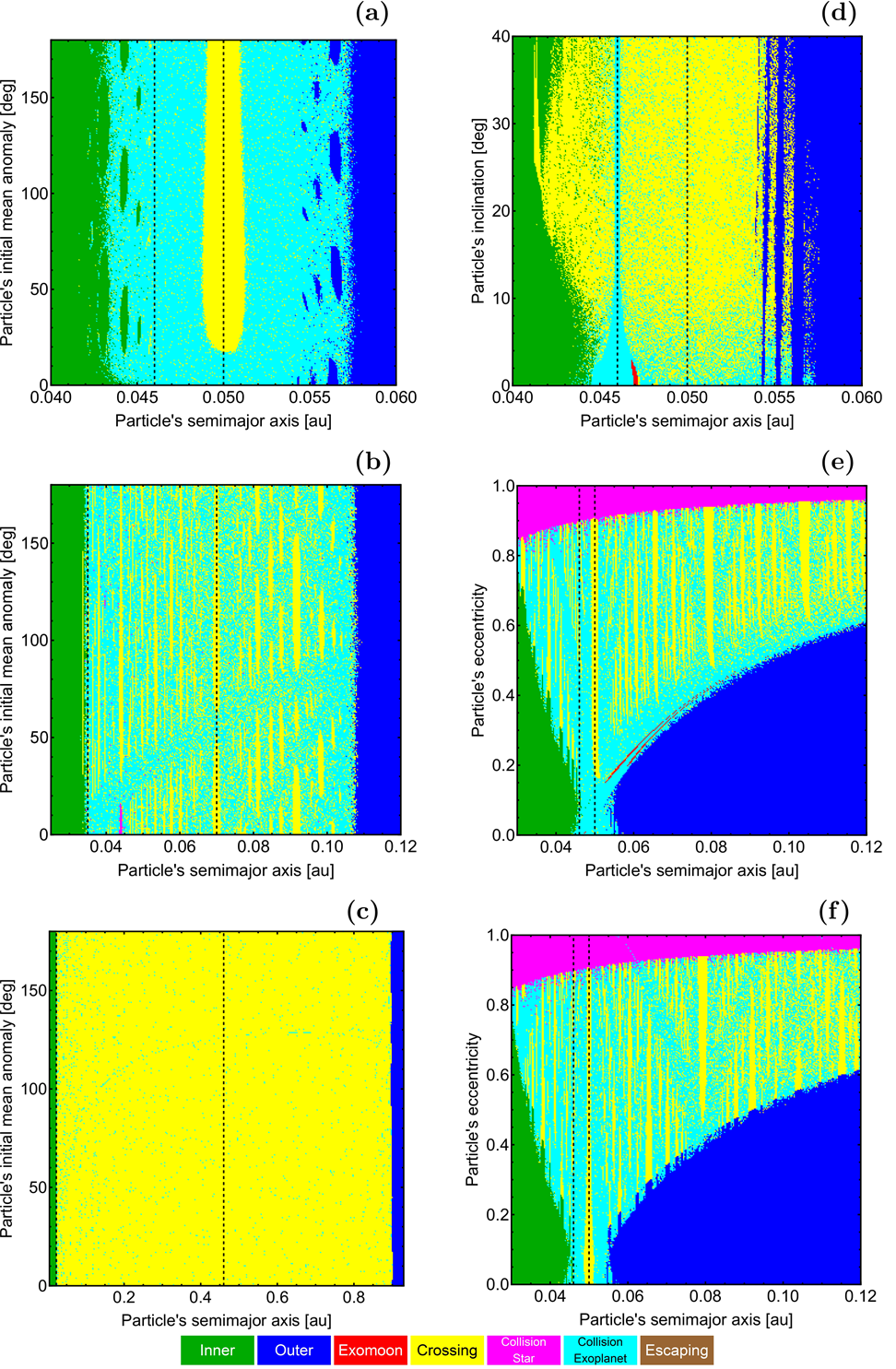}
\caption{
Particle orbit classification in the WD~J0914+1914 system assuming planet properties corresponding to Case III (panels {\bf a}, {\bf d}, {\bf e} and {\bf f}),  Case IV (panel {\bf b}) and Case V (panel {\bf c}). Despite the more complex behaviour exhibited by Neptune-like planets rather than Super-puffs, some results are similar: the planet does not significantly enhance pollution rates. However, the outer boundary of the gas disc becomes increasingly more nebulous as $e_{\rm pl}$ increases.
}
\label{neptune}
\end{figure*}

\begin{figure*}
\centerline{\bf \LARGE \underline{Eccentric Saturn-like planets}}
\includegraphics[width=14cm]{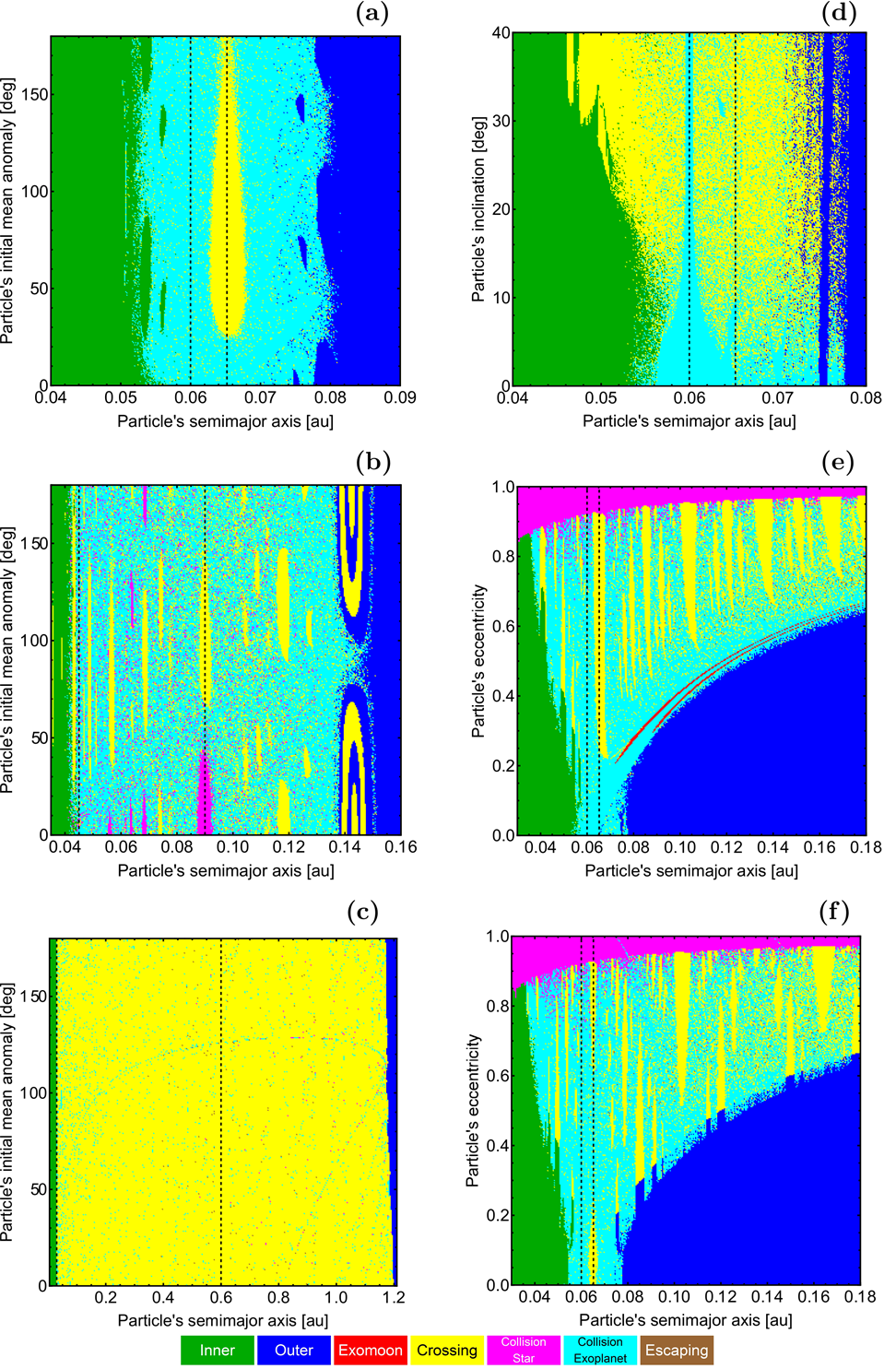}
\caption{
Particle orbit classification in the WD~J0914+1914 system assuming planet properties corresponding to Case VI (panels {\bf a}, {\bf d}, {\bf e} and {\bf f}),  Case VII (panel {\bf b}) and Case VIII (panel {\bf c}). The results are similar to Fig. \ref{neptune}, except that several of the panels here feature escape orbits, and panel {\bf b} shows a greater incidence of pollution orbits.
}
\label{saturn}
\end{figure*}

\begin{figure*}
\centerline{\bf \LARGE \underline{Varying planet mass for Saturn-like densities}}
\includegraphics[width=14cm]{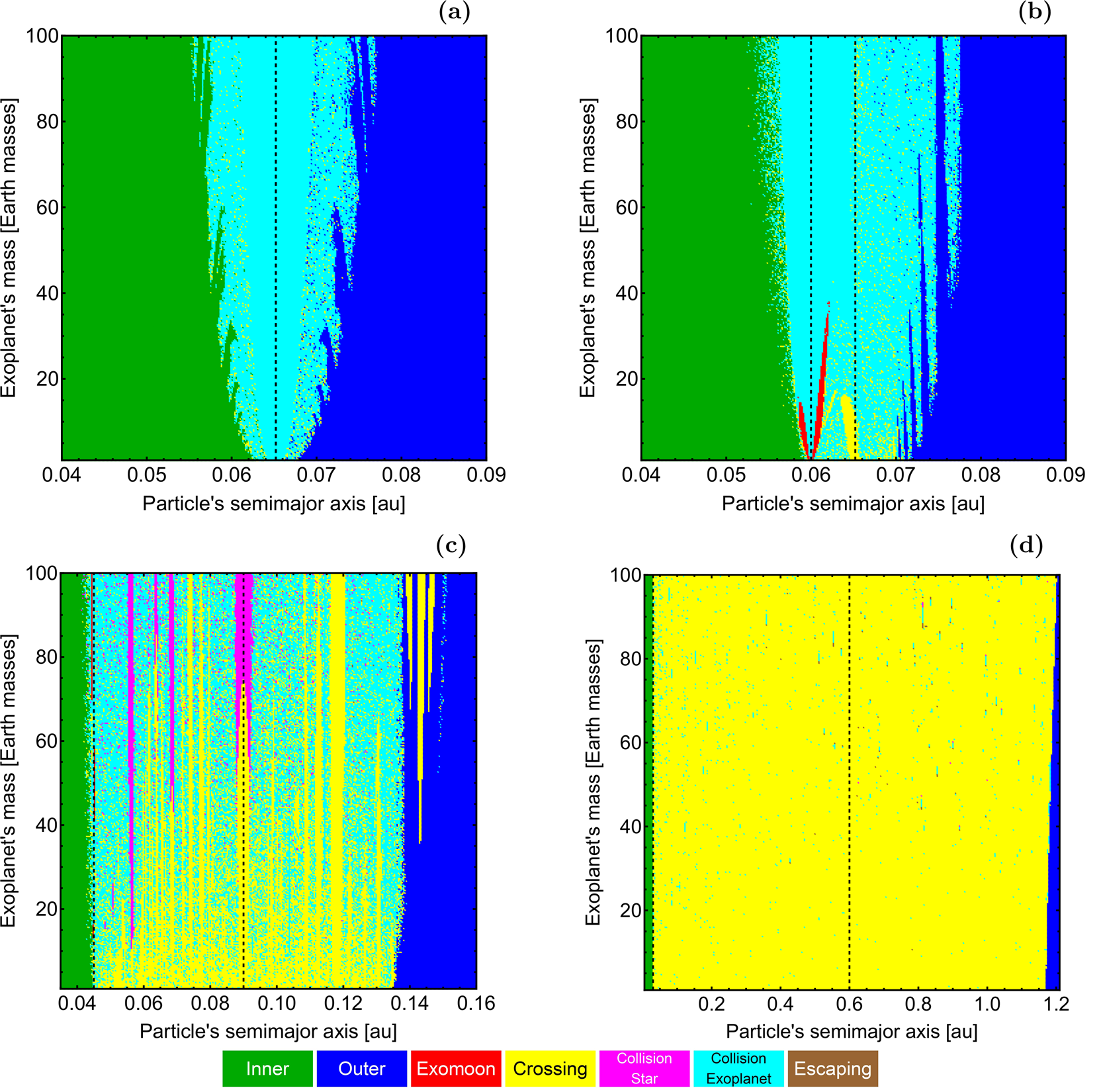}
\caption{
Particle orbit classification in the WD~J0914+1914 system, where the planet density is always assumed to be Saturn's density, but with $M_{\rm pl}$ varying between $1M_{\oplus}$ and $100M_{\oplus}$. In panels {\bf a, b, c} and {\bf d}, respectively, $e_{\rm pl} = \left\lbrace 0.00, 0.08, 0.50, 0.95 \right\rbrace$ and $a_{\rm pl} = \left\lbrace 0.0652, 0.0652, 0.09, 0.60 \right\rbrace$ au. Panel {\bf c} demonstrates that the number of polluting orbits increases with increasing $M_{\rm pl}$, but still occupies only a small region of parameter space. The outer boundary of the disc has a clear but relatively minor dependence on $M_{\rm pl}$.
}
\label{saturn2}
\end{figure*}

We present our results according to planet density. First we display the results for our Super-puffs (Cases I-II), followed by our Neptune analogues (Cases III-V) and then our Saturn analogues (Cases VI-VIII). The vertical black dotted lines on each plot correspond to the locations of both $q_{\rm pl}$ and $a_{\rm pl}$. In a particular plot, if a particle variable is not being varied then it takes on the following values: $e=i=\omega=\mathcal{M}_0 = 0^{\circ}$. Always, $\Omega = 0^{\circ}$.

\subsection{Super-puffs}

All of the Super-puff cases (Cases I-II) are shown in Fig. \ref{puff}. The top panels ({\bf a}: Case I; {\bf c}: Case II) illustrate particles with $e=i=0$, which may represent well-behaved disc particles. These panels demonstrate that a circular Super-puff carves out an instability gap at about 0.007 au for the $M_{\rm pl} = 1 M_{\oplus}$ planet and about 0.008 au for the $M_{\rm pl} = 4 M_{\oplus}$ planet. Hence, in no way this planet intrudes upon the assumed $0.0046-0.046$ au extent of the gas disc from \cite{ganetal2019}.

These upper panels also reveal U-shaped regions (solid cyan) within which the particle will always collide with the planet, regardless of $\mathcal{M}_0$. The transition from this region to one with crossing (yellow), inner (blue) or outer (green) orbits features different U-shaped regions containing a mixture of stable and unstable orbits. At the resolution of our integrations, these regions do not contain discernible resonances. In no case in the upper panels a particle impacts the star or escape the system.

Now consider the bottom panels of Fig. \ref{puff} where, for Case II, the particle eccentricity (panel {\bf b}) and inclination (panel {\bf d}) are varied. These particles more likely represent external dust which would veer close to the planet, rather than gas particles from a flat circular disc. Overall, when the particle's eccentricity or inclination deviates from zero, the types of orbit largely remain the same: collision with the planet, or stable circumstellar, circumbinary and crossing orbits. The only exceptions are a handful of circumplanetary ``moon'' orbits in panel {\bf d} -- but only for initial (circumstellar) values of $e \approx 0.1-0.2$ -- and polluting orbits, but only for $e \gtrsim 0.9$. Changing $i$ (panel {\bf c}) does not introduce any polluting orbits.

\subsection{Neptunes}

If the planet is, instead, a Neptune analogue, then it is both on an eccentric orbit and might reside within the disc (Cases III-V). Figure \ref{neptune} displays all of the Neptune-like cases.

In the left panels, $\mathcal{M}_0$ is varied for the low $e_{\rm pl}$ Case III (panel {\bf a}), moderate $e_{\rm pl}$ Case IV (panel {\bf b}) and high $e_{\rm pl}$ Case V ({\bf c}). All three panels are qualitatively different from the circular Super-puff case. Resonant structures are apparent except in panel {\bf c}, where just an arc of planet collision orbits is discernible. An increase in $e_{\rm pl}$ and a decrease in $q_{\rm pl}$ also naturally restricts the radial range of a stable gas disc. In fact, in panels {\bf b} and {\bf c}, the planet would be embedded within the disc.

What these plots demonstrate is that, for small to moderate planet eccentricities, the planet clears out external disc material. However, if the planet is still highly eccentric, then most external particles can survive. In all these cases, the boundary between circumstellar and collisional orbits is well represented by $q_{\rm pl}$ (leftmost dotted black line). Further, panel {\bf b} includes a small strip of polluting orbits at $a \approx 0.044$ au and $\mathcal{M}_0 \lesssim 15^{\circ}$, plus a smattering of other polluted orbits around the same semimajor axis.

The three right panels ({\bf d}-{\bf f}) of Fig. \ref{neptune} display Case III, but illustrate the phase space structure when $i$ is varied (panel {\bf d}) and when $e$ is varied (for $\omega = 0^{\circ}$ in panel {\bf e} and for $\omega = 180^{\circ}$ in panel {\bf f}). For all three panels, we see that the condition $a \approx q_{\rm pl}$ yields predominately (unstable) collisional orbits and the condition $a \approx a_{\rm pl}$ yields predominately (stable) crossing orbits. Panels {\bf e} and {\bf f} show abundant resonant structure, akin to that which was observed in \cite{zotetal2020a}.

Like in the circular planet case, here the resonant structure features a mix of planet collision orbits and stable crossing orbits. The most noticeable differences due to changing the value of $\omega$ is the lack of moon orbits and small string of stable circumbinary orbits at $e > 0.9$ in panel {\bf f}. None of the orbits in any of the three panels feature escape or collision with the star when $e < 0.75$.

In panel {\bf a} of Fig. \ref{neptune} we can identify several stability islands inside the collision area. These islands correspond to mean motion resonances of inner and outer trajectories. Moreover, in panels {\bf b}, {\bf e}, and {\bf f} of the same figure there are also present inside the collision basin numerous stability islands of crossing trajectories.

\subsection{Saturns}

A Saturn-analogue planet (Fig. \ref{saturn}) must reside on a different set of orbits (Cases VI-VIII) than a Neptune-like planet from Fig. \ref{neptune}. Consequently, and because Saturn analogues are 5.5 more massive than Neptune analogues, Saturn analogues carve out wider and more apparent resonant structures.

Nevertheless, Figs. \ref{neptune} and \ref{saturn} are qualitatively similar. Here, we highlight the differences. In panel {\bf a} (Case VI), the range of $\mathcal{M}_0$ for which the particle can reside on a stable crossing orbit is reduced. Panel {\bf b} (Case VII), showcases the most significant differences. This panel features several ``pollution strips'': phase space clusters of orbits that will pollute the white dwarf. One large strip occurs at $a_{\rm pl}$, and the others in between $q_{\rm pl}$ and $a_{\rm pl}$. These strips are centred around initial mean anomaly values of $\mathcal{M}_0 = \left\lbrace 0^{\circ}, 120^{\circ}, 180^{\circ} \right\rbrace$. In this panel, the gas disc outer boundary remains robust, despite some clusters of collisional orbits within 0.04 au. Panel {\bf c} (Case VIII), illustrates the first appearance of some escaping orbits, but not in any discernible pattern. For the right panels (all case VI), the main differences from those in Fig. \ref{neptune} are sharper resonant features and the presence of some escape orbits.

Figs. \ref{neptune} and \ref{saturn} displays snapshots of just two different planet masses. In order to probe more deeply into how the phase space structure changes as a result of $M_{\rm pl}$, we now present a series of plots where we vary the planet mass (Fig. \ref{saturn2}) but keep its density equivalent to that of Saturn's. We adopt four values of $e_{\rm pl}$ (0.00, 0.08, 0.50, 0.95) and corresponding values of $a_{\rm pl}$ (0.0652, 0.0652, 0.09, 0.60) au. Panel {\bf c} ($e_{\rm pl} = 0.50$), illustrates how the number of polluting orbits increases with $M_{\rm pl}$, corresponding to the trend seen in Figs. \ref{neptune} and \ref{saturn} (despite the difference in planet densities in those figures). Figure \ref{saturn2} also illustrates how increasing $M_{\rm pl}$ gradually alters the stability boundary of the outer limit of the gas disc, but never by more than 0.01 au.

\section{Discussion}
\label{disc}

Our results might help constrain and interpret upcoming observations, particularly with respect to the unknown mass and orbit of the planet. We have shown here that the outer boundary of the gas disc is primarily determined by both the orbital pericentre of the planet and its eccentricity. For low values of $e_{\rm pl}$, the unstable collisional orbits and stable circumstellar orbits are divided by the location which corresponds to the value of $q_{\rm pl}$. For high values of $e_{\rm pl}$, not only would the planet currently be embedded within the disc (see \citealt*{griver2019}), but also the outer boundary of the disc becomes fuzzier (containing pockets of stability).

Our results also clearly demonstrate the difficulty in polluting WD~J0914+1914 with rocky material, unless the progenitor is already on a highly eccentric orbit ($e > 0.75$). Particles on low- and moderate-eccentricity orbits will either remain stable or collide with the planet. Combined with the results of \cite{veras2020}, we claim that neither radiation nor gravitational forces in the nearby vicinity of the planet can effectively generate pollution, in line with the current observations.

Overall then, the scenario which is most consistent with the current observations is the one where the planet is a Super-puff on a circular or nearly circular orbit. More dense planets on more eccentric orbits would be consistent with a planet either embedded in the disc, or with a disc extent which is different than the one reported in \cite{ganetal2019}. Better observational constraints on the location or general structure of the gas disc will help distinguish these scenarios.

Some of the other dynamical features of our analysis also warrant discussion. In each part of the phase space, the orbit was declared to be stable or unstable. In none of the unstable cases did an escape occur, except for Saturn-analogue planets. This finding is sensible, and can be quantified with the Safronov number $\Theta$ \citep{safronov1972}

\[
\Theta = \frac{a_{\rm pl}}{R_{\rm pl}}
       \left(\frac{M_{\rm pl}}{M_{\star}} \right)
\]

\begin{equation}
\ \ \ \, = \left\lbrace
0.0023, 0.0059, 0.028, 0.039, 0.26, 0.085, 0.12, 0.79
\right\rbrace
\end{equation}

\noindent{}for, respectively, Cases I-VIII. The higher the Safronov number, the more likely escape can occur. For a given $a_{\rm pl}$, the highest Safronov numbers are obtained for Saturn-mass planets.

Also, some streaks on the plots might have more fundamental significance with regard to the three-body problem, but would need to be explored with longer integrations. The cyan streaks in panels {\bf c} of Figs. \ref{neptune} and \ref{saturn} could be reflective of unstable periodic orbits. Both \cite{antver2016} and \cite{antver2019} illustrated that, for both the circular and elliptic restricted three-body problems which include white dwarfs, unstable asteroids at high eccentricities ($e \gtrsim 0.95$) may reflect the locations of these periodic orbits. Alternatively, some periodic orbits can help to protect asteroids from polluting the white dwarf. For example, the blue streak of dots for $e > 0.9$ in panel {\bf f} of both Figs. \ref{neptune} and \ref{saturn} represent protected particles, perhaps from the vicinity of stable periodic orbits. Panels {\bf d} of Figs. \ref{neptune} and \ref{saturn} illustrate how inclination resonances are much harder to discern than eccentric resonances.

In a recent paper \citep{zotetal2020b} we performed a similar orbit classification, using the elliptic restricted three-body problem, for determining the character of motion of particles moving around Jupiter-like exoplanets. However, direct comparison between the results of the present work and the outcomes of the previous study \citep[or other general stability studies such as][]{ECM08} is not feasible. In particular, in \citet{zotetal2020b} we examined the nature of motion of the test particle, using initial conditions only near the vicinity of the secondary (exoplanet). Instead, here the initial conditions cover a much more extended area of the phase space around both the primary (star) and the secondary (exoplanet). Furthermore, the definition of the initial velocities of the test particle in both cases is not the same. Therefore, it is not possible to present a comparison between the results of the orbit classification, and doing so would be misleading given the significant differences between the two papers.

One could reasonably argue that because we consider a massless test particle (third body), we should use the elliptic version of the restricted three-body problem. However, we decided to use the general three-body problem (not the restricted one) for two main reasons: (i) the restricted version does not allow us to examine how important parameters of the system (such as the semi-major axis and inclination) affect the final states of the test particle and (ii) it is in our future plans to explore more complicated systems (e.g., with two stars and one exoplanet, one star and two exoplanets), for which the general three-body problem is the only option. Thus, we model the simplest scenario (one star, one exoplanet and a massless particle) first.

\section{Summary}
\label{sum}

We have analyzed the near-term stability of particles in the WD~J0914+1914 planetary system \citep{ganetal2019} in order to explore both the plausible boundaries for the gas disc in anticipation of upcoming observations and to understand why this white dwarf is not polluted with rocky debris. By integrating over a wide region of parameter space, we found that a Super-puff planet on a circular orbit best matches the current observations. However, if the planet is actually closer to the star than 0.07 au or is embedded within the disc, then the disc itself may feature significant substructure.

\section*{Data Availability Statements}

The data underlying this article will be shared on reasonable request to the corresponding author.

\section*{Acknowledgements}

We thank the anonymous referee for the careful reading of the manuscript as well as for all of the apt suggestions and comments, which have allowed us to improve both the quality and the clarity of the paper. DV gratefully acknowledges the support of the STFC via an Ernest Rutherford Fellowship (grant ST/P003850/1). LD acknowledges the support of the Agencia de Promoci\'{o}n Cient\'{i}fica, through PICT 201-0505 and PICT 2016-2635.

\label{lastpage}

\end{document}